\documentclass[runningheads,a4paper]{llncs}
\usepackage{amssymb}
\usepackage{graphicx}
\usepackage{comment}
\usepackage{amssymb}
\usepackage{amsmath}
\usepackage{paralist}
\usepackage{stmaryrd}
\usepackage{multirow}
\usepackage{helvet}
\usepackage{courier}
\usepackage{graphicx}
\usepackage{algorithmic}
\usepackage{graphics}
\usepackage{epsfig}
\usepackage{algorithm}
\usepackage{hyperref}
\usepackage{caption}
\usepackage{subfigure}
\usepackage{caption}

\usepackage{epstopdf}
\usepackage{xcolor}
\usepackage{tikz}
\usetikzlibrary{arrows}
\usetikzlibrary{decorations.markings}
\usetikzlibrary{shapes}
\usetikzlibrary{plotmarks}

\usepackage{multirow}
\usepackage{array,ragged2e}

\usepackage[compatibility=false]{caption}

\usepackage{color}
\usepackage{hhline}

\usepackage{pgfplots}

\usepackage{url}

\usepackage{tikz}
\usetikzlibrary{arrows,positioning,automata,decorations,fit,backgrounds,calc,shapes,snakes}
\usepgflibrary{shapes.geometric} % LATEX and plain TEX and pure pgf
\usetikzlibrary{shapes.geometric} % LATEX and plain TEX when using Tik Z
\usepgflibrary{decorations.pathmorphing} % LATEX and plain TEX and pure pgf
\usetikzlibrary{decorations.pathmorphing} % LATEX and plain TEX when using Tik Z

\urldef{\mailsa}\path|{zhihuiliu, zyfeng, xiaowangzhang, wangx, rgz}@tju.edu.cn|

\title{RORS: Enhanced Rule-based OWL Reasoning on Spark}

\titlerunning{Enhanced Rule-based OWL Reasoning on Spark}
\authorrunning{Z. Liu, Z. Feng, X. Zhang, X. Wang, \& G. Rao}

\author{Zhihui Liu \and Zhiyong Feng \and Xiaowang Zhang \and Xin Wang \and Guozheng Rao}
\institute{School of Computer Science and Technology, Tianjin University, China\\
Tianjin Key Laboratory of Cognitive Computing and Application, Tianjin, China\\
\mailsa}
\begin{document}

\maketitle

\begin{abstract}
The rule-based OWL reasoning is to compute the deductive closure of an ontology by applying RDF/RDFS and OWL entailment rules. The performance of the rule-based OWL reasoning is often  sensitive to the rule execution order. In this paper, we present an approach to enhancing the performance of the rule-based OWL reasoning on Spark based on a locally optimal executable strategy. Firstly, we divide all rules (27 in total) into four main classes, namely, \emph{SPO rules} (5 rules), \emph{type rules} (7 rules), \emph{sameAs rules} (7 rules), and \emph{schema rules} (8 rules) since, as we investigated, those triples corresponding to the first three classes of rules are overwhelming (e.g., over 99\% in the LUBM dataset) in our practical world. Secondly, based on the interdependence among those entailment rules in each class, we pick out an optimal rule executable order of each class and then combine them into a new rule execution order of all rules. Finally, we implement the new rule execution order on Spark in a prototype called RORS. The experimental results show that the running time of RORS is improved by about 30\% as compared to Kim \& Park's algorithm (2015) using the LUBM200 (27.6 million triples).
\end{abstract}

\section{Introduction}\label{sec:intro}
The Web Ontology Language \cite{owl} (OWL) is a semantic web language designed to represent rich and complex knowledge about
things, groups of things, and relations between things. There are mainly two modeling paradigms for the semantic web. The first paradigm
is based on the notion of the classical logics, such as the description logics algorithms \cite{Badder2001tableau} on which the OWL is based. The
other paradigm is based on the Datalog paradigm. A subset of the OWL semantics is transformed into rules that are used by a rule engine in order
to infer implicit knowledge. This paper focuses on the second paradigm, that is, a rule-based OWL reasoning using OWL-Horst \cite{chang2011fuzzy} rules.

Owing to the explosion of the semantic data, the number of RDF triples in large public knowledge bases, e.g., DBpedia has increased to billions \cite{rong2015cichlid}. Therefore, to improve the performance of OWL reasoning becomes a core problem. The traditional single-node approaches are no longer viable for such large scale data. Some existing ontology reasoning systems are based on MapReduce framework \cite{maeda2014mapreduce,urbani2009scalable,urbani2010owl,wu2013distributed}. The OWL reasoning in \cite{wu2013distributed} and \cite{maeda2014mapreduce} perform reasoning over MapReduce with rule execution mechanism. However, the MapReduce-based approaches are not very efficient due to the data communication between memory and disk. To further improve the performance of reasoning, some researchers have implemented the OWL reasoning on Spark, which is an in-memory and distributed cluster computing framework \cite{spark}.

Recently, Cichlid\cite{rong2015cichlid} has greatly improved the performance of OWL reasoning on Spark as compared to the state-of-the-art distributed reasoning systems, but it only considers parts of OWL rules and does't analyze the interdependence of rules. Reasoning based on OWL-Horst rules can infer many more implicit information. And different rule execution strategy will influence the reasoning performance. For instance, let S be the set of triples of an ontology, where $S$=\{$\langle A$ \emph{subClassOf} $B\rangle$, $\langle B$ \emph{subClassOf} $C\rangle$\}. R1 and O11 are the two rules of OWL-Horst, where $R1$=\{$\langle C$ \emph{rdfs:subClassOf} $C1\rangle$, $\langle C1$ \emph{rdfs:subClassOf} $C2\rangle$\ $\Rightarrow$ $\langle C$ \emph{rdfs:subClassOf} $C2\rangle$\} and $O11$=\{$\langle V$ \emph{owl:equivalentClass} $W\rangle$ $\Rightarrow$ $\langle V$ \emph{rdfs:subClassOf} $W\rangle$ \}. By implementing the R1 entailment rule for subclass closure, we will get that $S$=\{$\langle A$ \emph{subClassOf} $B\rangle$, $\langle B$ \emph{subClassOf} $C\rangle$, $\langle A$ \emph{subClassOf} $C\rangle$\}. If the O12 entailment rule for equivalent class is executed before the R1, S will contain more new triples and reduce the iterative operation. Therefore, it is desired to optimize reasoning by adjust the rule order. Although Kim \& Park \cite{je2015scalable} has also implemented parallel reasoning algorithms with an executable rule order, but lacked the evidence to prove that the strategy is optimal.

To find the optimal executable strategy, we use the depth-first algorithm to get all possible executable strategies,
which are based on the dependency of rules. There are 259367372 possible strategies among the 27 rules in Table \ref{tab:rules}. Due to the very large number of strategies, it is challenge to find the optimal strategy by test every strategy.

In this paper, we present an approach to enhancing the performance of the rule-based OWL reasoning based on a locally optimal executable strategy and implement the new rule execution strategy on Spark in a prototype called RORS. The major contributions and novelties of our work are summarized as follows:
\begin{compactitem}
\item We analyze the characteristic of dataset and divide the dataset into three classes: SPO triples, sameAs triples and type triples with analysing the proportion of the three classes respectively in the dataset. According to the data partition, we divide the OWL-Horst rules into four classes.
\item We respectively analyze the rule interdependence of each class, and find the optimal executable strategies.
\item Based on the locally optimal strategies, we pick out an optimal rule execution order of each class and then combine them into a new rule execution strategy of all rules and implement the new rule execution strategy on Spark.
\end{compactitem}

The rest of this paper is organized as follows. Section \ref{sec:pre} gives an brief introduce to preliminary knowledge about OWL and Spark. Section \ref{sec:Rda} presents our locally optimal strategies. Section \ref{sec:rea} implements our proposed strategy on Spark and Section \ref{sec:eva} evaluates the experiment on the LUBM dataset. In Section \ref{sec:dis}, we discuss related works and summarize this paper.

\section{Preliminaries}\label{sec:pre}
In this section, we briefly recall the ontology language OWL and the framework Spark, largely
following the excellent expositions \cite{owl,spark}.
\paragraph{\bf OWL}
An ontology is a formal naming and definition of the types, properties, and interrelationships of the entities that really or fundamentally
exist for a particular domain of discourse. Ontology is part of the W3C standards stacks for the semantic web. The language OWL \cite{owl} is a family of knowledge representation languages for authoring ontologies. There are three variants of OWL, with different levels of expressiveness. These are OWL Lite, OWL DL and OWL Full (ordered by increasing expressiveness). Each of these sublanguages is
a syntactic extension of its simpler predecessor. OWL DL is designed to
preserve some compatibility with RDF Schema (or RDFS). However, OWL Full is
undecidable, so no reasoning software is able to perform complete reasoning
for it. OWL DL designed to provide the maximum expressiveness possible while
retaining computational completeness, decidability, and the availability of
practical reasoning algorithms. OWL Lite was originally intended to support
those users primarily needing a classification hierarchy and simple
constraints. The three languages are one subset of the other.

\paragraph{\bf Spark: Distributed computing framework}
Spark \cite{spark} is an open source cluster computing framework, which is developed at
the University of California, Berkeley's AMPLab. One of the main features is
the in-memory parallel computing model which all data will be loaded into
the memory. Spark provides programmers with an application programming
interface centered on a data structure called the resilient distributed
dataset \cite{zaharia2012resilient} (RDD), a read-only multiset of data items distributed over a cluster
of machines, that is maintained in a fault-tolerant
way. Each RDD will be divided into multiple
partitions that exist on different computing nodes. And it provides a
variety of operations to transform one RDD into another RDD. There are two
kinds of operations. Transformations are lazy operations that define a new
RDD (e.g., \emph{map}, \emph{filter}, and \emph{join}), while actions launch a computation to
return a value to the program or write data to external storage, such as
\emph{collect}, \emph{count}, \emph{saveAsTextFile}, etc. RDD achieves fault tolerance through a notion of lineage based on logging the transformations, if a partition of an
RDD is lost, the RDD has enough information about how it was derived from
other RDDs to recompute just that partition. More details about Spark please
see the official web site \url{http://spark.apache.org/}.

\section{Locally optimal strategy}\label{sec:Rda}
In this section, we propose a locally optimal strategy based on the dependency among the rules.
\begin{table}[H]
  \centering
  \caption{OWL-Horst rules}\label{tab:rules}
  \vspace*{3mm}
  \begin{tabular}{|c|c|c|}
    \hline
    Rule ID & Condition & Consequence \\
    \hline
    \multirow{2}{*}{R1} & c rdfs:subClassOf c1 & \multirow{2}{*}{c rdfs:subClassOf c2} \\
       & c1 rdfs:subClassOf c2 & \\
    \hline
    \multirow{2}{*}{R2} & p rdfs:subPropertyOf p1 & \multirow{2}{*}{p rdfs:subPropertyOf p2}\\
       & p1 rdfs:subPropertyOf p2 & \\
    \hline
    R3 & s p o , p rdfs:subPropertyOf p1 & s p1 o\\
    \hline
    R4 & s rdfs:domain x , u s y & u rdf:type x\\
    \hline
    R5 & p rdfs:range o , s p v & v rdf:type o\\
    \hline
    R6 & c rdfs:subClassOf c1 , v rdf:type c & v rdf:type c1\\
    \hline
      \multirow{2}{*}{O1} & p rdf:type owl:FunctionalProperty  & \multirow{2}{*}{v owl:sameAs w}\\
       & u p v , u p w & \\
    \hline
     \multirow{2}{*}{O2} & p rdf:type owl:InverseFunctionalProperty  & \multirow{2}{*}{v owl:sameAs w}\\
       & v p u , w p u & \\
     \hline
     \multirow{2}{*}{O3} & p rdf:type owl:SymmetricProperty & \multirow{2}{*}{u p v}\\
       & v p u & \\
     \hline
     \multirow{2}{*}{O4} & p rdf:type owl:TransitiveProperty  & \multirow{2}{*}{u p v}\\
       & u p w , w p v & \\
     \hline
     O5 & v owl:sameAs w & w owl:sameAs v\\
     \hline
     \multirow{2}{*}{O6} & v owl:sameAs w & \multirow{2}{*}{v owl:sameAs u}\\
       & w owl:sameAs u & \\
     \hline
     O7a & p owl:inverseOf q, v p w & w q v\\
     \hline
     O7b & p owl:inverseOf q, v q w & w p v\\
     \hline
      \multirow{2}{*}{O8} & v rdf:type owl:Class & \multirow{2}{*}{v rdfs:subClassOf w}\\
       & v owl:sameAs w & \\
      \hline
      \multirow{2}{*}{O9} & p rdf:type owl:Property & \multirow{2}{*}{p rdf:subPropertyOf q}\\
       & p owl:sameAS q & \\
      \hline
      \multirow{2}{*}{O10} & u p v, u owl:sameAs x  & \multirow{2}{*}{x p y}\\
       & v owl:sameAs y & \\
      \hline
      O11a & v owl:equivalentClass w & v rdfs:subClassOf w\\
      \hline
      O11b & v owl:equivalentClass w & w rdfs:subClassOf v\\
      \hline
      \multirow{2}{*}{O11c} & v owl:subClassOf w & \multirow{2}{*}{v rdfs:equivalentClass w}\\
        & w owl:subClassOf v & \\
      \hline
      O12a & v owl:equivalentProperty w & v rdfs:subPropertyOf w\\
      \hline
      O12b & v owl:equivalentProperty w & w rdfs:subPropertyOf v\\
      \hline
      \multirow{2}{*}{O12c} & v owl:subPropertyOf w & \multirow{2}{*}{v rdfs:equivalentProperty w}\\
        & w owl:subPropertyOf v & \\
      \hline
         \multirow{2}{*}{O13} & v owl:hasValue w  & \multirow{2}{*}{u rdf:type v}\\
       & v owl:onProperty p, u p v & \\
      \hline
           \multirow{3}{*}{O14} & v owl:hasValue w  & \multirow{3}{*}{u p v}\\
       & v owl:onProperty p & \\
       & u rdf:type v & \\
       \hline
          \multirow{2}{*}{O15} & v owl:someValuesFrom w  & \multirow{2}{*}{u rdf:type v}\\
       & v owl:onProperty p & \\
       & u p x,  x rdf:type w & \\
       \hline
          \multirow{2}{*}{O16} & v owl:allValuesFrom w  & \multirow{2}{*}{x rdf:type w}\\
       & v owl:onProperty p & \\
       & u rdf:type v, u p x & \\
       \hline
  \end{tabular}
\end{table}
\subsection{Dependance of rules}
OWL-Horst \cite{chang2011fuzzy} has a powerful expression and reasoning ability. In this paper, the OWL reasoning is based on OWL-Horst rules shown in Table \ref{tab:rules}. There exists interdependency among the rules. For example, the output of O13 can be the condition of O14,
whereas O14 affects O3, O4, O7a, O7b and R3. And more than one rule can be the input of another rule. The output of O3, O4, and R3
affects O13, O15 and O16. If the input to a rule R$_i$ depends on the output of another rule R$_j$, then the rule R$_i$ is dependent on R$_j$. The rule R$_j$ should be executed before the rule R$_i$. A rule dependency graph can be constructed based on the interdependency among the rules. Each vertex represents a rule and an outgoing edge between vertex v$_{i}$ and v$_{j}$ represents the dependency of vertex v$_{i}$ on v$_{j}$. The dependency graph of SPO rules is shown in the Figure \ref{SPO}.
Based on the dependency graph, there are 259367372 executable strategies among the 27 rules that adjacent rules satisfy dependency, and each rule is executed only once. And the longest strategies contain 22 rules that cannot fully sequentially execute all rules. The number is so huge that it is challenge to test every strategy. It is known that reasoning has close relationship with the characteristic of datasets. Therefore, we prefer to make a analysis of LUBM dataset.

\subsection{Locally optimal strategy}
LUBM \cite{guo2005lubm} is a widely used standard benchmark for evaluating the performance of ontology reasoning. We can use data generator to
generate dataset with different size. Data generator is carried out by the Univ-Bench Artificial data generator (UBA), a tool developed for the
benchmark. We firstly divide the dataset into three classes as follows:
\begin{compactitem}
  \item Triples whose predicate is \emph{rdf:type}. We call this class as type.
  \item Triples whose predicate is \emph{owl:sameAs}. We call this class as \emph{sameAs} .
  \item The remainder is classified as a class. We call this class as SPO.
\end{compactitem}
The UBA uses specific rules to generate data so that all datasets generated by it have same data characteristic. So we use the
dataset LUBM-50 as the sample and analyze the proportion of each class. The result is listed in Table \ref{lubm:pro}.
\vspace{-1cm}
\begin{table}[h]
  \centering
  \caption{The proportion of each type in LUBM-50}\label{lubm:pro}
  \vspace{3mm}
  \begin{tabular}{|c|c|c|c|}
    \hline
    Dataset & type & sameAs & SPO \\
    \hline
    LUBM-50 & 20.055\% & 0 & 79.945\% \\
    \hline
  \end{tabular}
\end{table}
\vspace*{-6mm}

From the Table \ref{lubm:pro}, we can see that the SPO triples of LUBM\_50 account for absolute proportion, about 80\% of the total. The type triples are in the second place, but the number of \emph{sameAs} triples is zero. Therefore, the OWL reasoning should focus on the reasoning of SPO triples and type triples. The \emph{sameAs} triples are simply handled. Based on the statistic, we divide the OWL-Horst rules into four classes as follow:
\begin{compactitem}
  \item Rules whose condition or consequence has triples of type class, including R4, R5, R6, O13, O14, O15, O16. These rules are used to infer implicit type data. The dependency graph of type rules is shown in Figure \ref{type}.
  \item Rules whose condition or consequence has triples of \emph{sameAs} class, including O1, O2, O5, O6, O8, O9, O10. There are certain rules for ontology merging(O8 and O9) \cite{je2015scalable}, so we also exclude these rules from our reasoner. The dependency graph of sameAs rules is shown in Figure \ref{sameAs}.
  \item Rules whose condition or consequence has triples of SPO class, including R3, O3, O4, O7a, O7b. The dependency graph of SPO rules is shown in Figure \ref{SPO}, which the rule O7a and O7b are classified as O7.
  \item The remainder is classified as a class, including R1, R2, O11a, O11b, O11c, O12a, O12b, O12c. The dependency graph of schema rules is shown in Figure \ref{schema}.
\end{compactitem}

\begin{figure}[H]
\setlength{\abovecaptionskip}{0.25cm}
\setlength{\belowcaptionskip}{-0.65cm}
  \centering
  \begin{minipage}[t]{0.5\linewidth}
     \begin{tikzpicture}[line width=0.7pt]
        \node[shape=circle, fill=gray!40] (R3) at (2, 1.8) {R3};
        \node[shape=circle, fill=gray!40] (O7) at (2, 3) {O7};
        \node[shape=circle, fill=gray!40] (O3) at (1, 1) {O3};
        \node[shape=circle, fill=gray!40] (O4) at (4, 1) {O4};
        \draw[<->] (R3)--(O7);
        \draw[<->] (R3)--(O3);
        \draw[<->] (R3)--(O4);
        \draw[<->] (O3)--(O4);
        \draw[<->] (O7)--(O4);
    \end{tikzpicture}
  \caption{SPO rules dependency graph}\label{SPO}
  \end{minipage}%
  \begin{minipage}[t]{0.5\linewidth}
      \begin{tikzpicture}[scale=0.7, line width=0.7pt]
        \node[shape=rectangle, fill=gray!40] (a) at (2, 1) {R4 R5 O13};
        \node[shape=circle, fill=gray!40] (b) at (-1, 0) {R6};
        \node[shape=circle, fill=gray!40] (c) at (5, 0) {O14};
        \node[shape=rectangle, fill=gray!40] (d) at (2, 3) {O15 O16};
        \draw[->] (a)--(b);
        \draw[->] (a)--(c);
        \draw[->] (a)--(d);
        \draw[->] (b)--(c);
        \draw[->] (d)--(c);
        \draw[<->] (b)--(d);
      \end{tikzpicture}
     \caption{type rules dependency graph}\label{type}
  \end{minipage}
\end{figure}
\vspace*{-0.5cm}
\begin{figure}[H]
\setlength{\abovecaptionskip}{0.25cm}
\setlength{\belowcaptionskip}{-0.65cm}
  \centering
  \begin{minipage}[t]{0.5\linewidth}
     \begin{tikzpicture}[line width=0.7pt]
        \node[shape=rectangle, fill=gray!40] (a) at (2, 0) {O1 O2};
        \node[shape=circle, fill=gray!40] (b) at (2, -1) {O10};
        \node[shape=circle, fill=gray!40] (c) at (0, 0.5) {O5};
        \node[shape=circle, fill=gray!40] (d) at (4, 0.5) {O6};
        \draw[->] (a)--(b);
        \draw[->] (a)--(c);
        \draw[->] (a)--(d);
        \draw[->] (c)--(b);
        \draw[->] (d)--(b);
        \draw[<->] (c)--(d);
     \end{tikzpicture}
     \caption{sameAs rules dependency graph}\label{sameAs}
  \end{minipage}%
  \begin{minipage}[t]{0.5\linewidth}
    \begin{tikzpicture}[line width=0.7pt]
        \node[shape=rectangle, fill=gray!40] (a) at (-2, 1) {O11a O11b};
        \node[shape=rectangle, fill=gray!40] (b) at (0, 1) {R1};
        \node[shape=rectangle, fill=gray!40] (c) at (-1, 0) {O11c};
        \node[shape=rectangle, fill=gray!40] (d) at (1.5, 1) {O12a O12b};
        \node[shape=rectangle, fill=gray!40] (e) at (3, 1) {R2};
        \node[shape=rectangle, fill=gray!40] (f) at (2, 0) {O12c};
        \draw[->] (a)--(b);
        \draw[<->] (a)--(c);
        \draw[->] (b)--(c);
        \draw[->] (d)--(e);
        \draw[<->] (d)--(f);
        \draw[->] (e)--(f);
    \end{tikzpicture}
    \caption{schema rules dependency graph}\label{schema}
  \end{minipage}%
\end{figure}
%\vspace*{-0.5cm}

For each class, based on the dependency of rules, we use \emph{depth-first-search} (DFS) algorithm to find all possible executable orders among the
rules and acquire the optimal executable orders. Through the experiment, there is a large number of orders for each class, but we should chose
the longest orders that contain rules as many as possible. The optimal executable orders for each class are listed in Table \ref{paths}.
\vspace{-1cm}
\begin{table}[H]
    \centering
    \caption{The optimal strategies of each class}\label{paths}
    \vspace{3mm}
    \begin{tabular}{|c|c|}
      \hline
      SPO rules & type rules \\
       & R4 $\to$ R6 $\to$ O14 $\to$ O13 $\to$ O15 $\to$ O16  \\
      O3 $\to$ R3 $\to$ O7 $\to$ O4 & R4 $\to$ R6 $\to$ O14 $\to$ O13 $\to$ O16 $\to$ O15  \\
      O7 $\to$ R3 $\to$ O3 $\to$ O4 & R5 $\to$ R6 $\to$ O14 $\to$ O13 $\to$ O15 $\to$ O16  \\
                                    & R5 $\to$ R6 $\to$ O14 $\to$ O13 $\to$ O16 $\to$ O15 \\
      \hline
      schema rules & sameAs rules \\
      O11a, O11b $\to$ R1 $\to$ O11c & O1 $\to$ O10 $\to$ O2 $\to$ O6 $\to$ O5 \\
      O12a, O12b $\to$ R2 $\to$ O12c & O2 $\to$ O10 $\to$ O1 $\to$ O6 $\to$ O5 \\
      \hline
    \end{tabular}
\end{table}

The symmetric rule O3 and the \emph{inverseOf} rule O7 are
not affected by each other because a property can not simultaneously involve these two feathers \cite{je2015scalable}. So O7 and O3 don't have dependence. The
transitive rule O4 is used to compute the SPO triples closure and should be applied at last in SPO rules before generating more new triples. The
second part is connected by type triples. We find that each path lacks R4 or R5 because there is no type triples in their condition, so they
should be classified to same kind of rule. The R4 and R5 can be executed in any order. The third part works with schema triples. Because the
O11a and O11b produce similar triples with the only different order of subject and object, so there is no order between O11a and O11b. In the
condition of O1 and O2, owing to the different property of the predict, the paths are equivalent when exchanging the order between O1 and O2.
The O5 does not produce new triples actually and should be placed at last.

\section{Distributed rule-based OWL reasoning on Spark}\label{sec:rea}
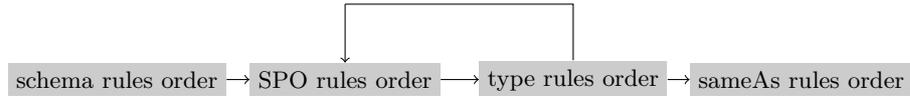
\begin{figure}
  \centering
  \begin{tikzpicture}
  \node[shape=rectangle, fill=gray!40] (schema) at (2, 3) {schema rules order};
  \node[shape=rectangle, fill=gray!40] (SPO) at (5, 3) {SPO rules order};
  \node[shape=rectangle, fill=gray!40] (type) at (8, 3) {type rules order};
  \node[shape=rectangle, fill=gray!40] (sameAs) at (11, 3) {sameAs rules order};
  \draw[->] (schema)--(SPO);
  \draw[->] (SPO)--(type);
  \draw[->] (type)--(sameAs);
  \draw (type.north) -- (8, 4) -- (5, 4);
  \draw[->] (5, 4) -- (SPO.north);
  \end{tikzpicture}
  \caption{The reasoning strategy}\label{strategy}
\end{figure}

There are two kinds of triples in OWL reasoning. The schema triples are OWL rule schema which provides basic elements for the description of ontology while the instance triples are actual statement in ontology, including SPO triples, type triples and sameAs triples. After proposing the locally optimal orders for each class, we design the overall strategy of the OWL reasoning shown in Figure \ref{strategy}.

The reasoning of instance triples will use the output of schema triples, so the schema reasoning should be executed firstly. There are several orders for each class of instance triples. We can chose any one order and then combine an new executable strategy. In this paper, we select first order for each class to perform reasoning. The general workflow of the parallel OWL reasoning is described in Algorithm \ref{alg1}.
\begin{algorithm}
{\fontsize{8pt}{8pt}\selectfont
\caption{OWL reasoning algorithm on Spark}
\label{alg1}
\begin{algorithmic}[1]
    \REQUIRE triples, OWL\;Horst\; rule\; set
    \ENSURE result
    \STATE schema\_derived=triples.apply(schema\_rules)
    \STATE val \ flag=true
    \WHILE{flag}
        \STATE SPO\_derived=triples.apply(SOP\_rules)
    \IF{SPO\_derived!=null}
        \STATE triples=triples.\textbf{union}(SPO\_derived)
    \ENDIF
    \STATE type\_derived=triples.apply(type\_rules)
    \IF{type\_derived!=null}
        \STATE triples=triples.\textbf{union}(type\_derived)
    \ENDIF
    \IF{SPO\_derived==null \&\& type\_derived==null}
        \STATE flag=false
    \ENDIF
    \ENDWHILE
    \STATE sameAs\_derived=triples.apply(sameAs\_rules)
    \STATE triples=triples.\textbf{union}(sameAs\_derived)
    \RETURN triples
\end{algorithmic}
}
\end{algorithm}

\paragraph{\bf Optimize join operation}
It is inevitable that there exists a lot of join operations on triples, especially the multiple triples in the condition, such as O14.
The join operation can greatly influence the performance of reasoning. Generally, the schema triples of an ontology is small even if the ontology is very large \cite{je2015scalable}.
Spark provides the broadcast variables which can transform local data to all available computing nodes and keep them cached on each machine. In our
method, we adopt the mechanism \cite{rong2015cichlid} dealing with join operation, which broadcasted the schema triples to every computing node
before the join operation. For the rules containing multi-join operation, the join operation
inside the schema triples can firstly be executed in memory locally. The instance triples will be divided into many partitions and then execute the
join operation between schema triples and instance triples in parallel. Here we take the O16 as an example to describe the operation. The execution
procedure is described in Algorithm \ref{alg2}.

\begin{algorithm}
{\fontsize{8pt}{8pt}\selectfont
\caption{Optimized the join operation of rule O16 using broadcast variables}
\label{alg2}
\begin{algorithmic}[1]
    \REQUIRE triples
    \ENSURE results
    \STATE val triples=sc.\textbf{textFile}(``hdfs://...'')
    \STATE val op=triples.\textbf{filter}(t$\Rightarrow$t.\_2.equals(``owl:onProperty'')).\textbf{map}(t$\Rightarrow$(t.\_1, t.\_3))
    \STATE \quad .\textbf{collect}.toMap
    \STATE val opBroadcast=sc.\textbf{broadcast}(op)
    \STATE val av=triples.\textbf{filter}(t$\Rightarrow$t.\_2.equals(``owl:allValuesFrom'')).\textbf{map}(t$\Rightarrow$(t.\_1, t.\_3))
    \STATE \quad .\textbf{collect}.toMap
    \STATE val avBroadcast=sc.\textbf{broadcast}(av)
    \STATE val ins=triples.\textbf{filter}(t$\Rightarrow$av.value.contains(t.\_1) \&\& op.value.contains(t.\_1))
    \STATE \quad .\textbf{map}(t$\Rightarrow$((t.\_1, op.value(t.\_2)), av.value(t.\_2)))
    \STATE val typ=triples.\textbf{filter}(t$\Rightarrow$op2.contains(t.\_2) \&\& av.value.contains(op2(t.\_2)))
    \STATE \quad .\textbf{map}(t $\Rightarrow$((t.\_1, t.\_2), t.\_3))
    \STATE val results=ins.\textbf{join}(type.\textbf{map}(t$\Rightarrow$(t.\_2.\_2, t.\_2.\_1)))
    \RETURN results
\end{algorithmic}
}
\end{algorithm}

\begin{algorithm}
{\fontsize{8pt}{8pt}\selectfont
\caption{transitive closure algorithm}
\label{alg3}
\begin{algorithmic}[1]
    \REQUIRE transitive triples
    \ENSURE results
        \STATE  var flag=1L
        \STATE  var p=triples
        \STATE  var q=p
        \STATE  var l=q.\textbf{map}(t $\Rightarrow$((t.\_2, t.\_3), t.\_1)).\textbf{partitionBy}(partitioner)
    \WHILE{flag!= 0}
        \STATE  val r=q.\textbf{map}(t $\Rightarrow$((t.\_2, t.\_1), t.\_3)).\textbf{partitionBy}(partitioner)
        \STATE  val q1=l.\textbf{join}(r).\textbf{map}(t $\Rightarrow$(t.\_2.\_1, t.\_1.\_1, t.\_2.\_2))
        \STATE  q=q1.\textbf{subtract}(p, parallism).\textbf{persist}(storagelevel)
        \STATE  flag=q.\textbf{count}()
    \IF{flag!=0}
        \STATE  l= q.\textbf{map}(t $\Rightarrow$((t.\_2, t.\_3), t.\_1)).\textbf{partitionBy}(partitioner)
        \STATE val s=p.\textbf{map}(t $\Rightarrow$((t.\_2, t.\_1), t.\_3)).\textbf{partitionBy}(partitioner)
        \STATE val p1=s.\textbf{join}(l).\textbf{map}(t $\Rightarrow$(t.\_2.\_2, t.\_1.\_1, t.\_2.\_1)).\textbf{persist}(storagelevel)
        \STATE p=p1.\textbf{union}(q).\textbf{union}(p)
    \ENDIF
    \ENDWHILE
    \RETURN p
\end{algorithmic}
}
\end{algorithm}

\paragraph{\bf Reasoning for \emph{owl:sameAs} property}
The sameAs rules derived too many triples and most of the output data is valueless in practical applications \cite{urbani2010owl}. To improve the performance,
many existing methods prefer to build a sameAsTable \cite{urbani2010owl}, \cite{rong2015cichlid}, which design a hash-code function and compare the hash
value. However, dealing with \emph{owl:sameAs} property often consume too much time. Based on the character of dataset, we do not adopt
the above method but simplify the reasoning process. The O6 can also be dealt with the transitive closure Algorithm \ref{alg3}.

\section{EVALUATION}\label{sec:eva}
In this section, we conduct a series of experiments to compare the performance of the proposed approach with the KP \cite{je2015scalable} and
Cichlid \cite{rong2015cichlid} under the same environment.

\subsection{Experiment and Dataset}
We set up a cluster with one master and four worker nodes. Each node has 48 Xeon E5\_4607 2.20GHz processors, 64GB memory and 10TB 7200 RPM SATA hard
disk. The nodes are connected with Gigabit Ethernet. All the nodes run on 64-bit Ubuntu 12.04 LTS operating system and Ext3 file system. The version
of the Spark is 1.0.2. And the corresponding Hadoop v2.2 with Java1.7 is installed on this cluster. The Spark has special requirement for the version
of Hadoop. Besides, the version of the Scala is 2.10.6. We use the synthetic benchmarks. LUBM \cite{guo2005lubm} is a widely used standard benchmark
for semantic program. Due to the limitation of hardware, we use the data generator UBA to generate 5 sets of data with different universities: LUBM\_10, LUBM\_50,
LUBM\_100, LUBM\_150 and LUBM\_200 in our experiment. The number of triples in each data set is shown in Table \ref{lubm} .
\begin{table}[H]
  \small
  \centering
  \caption{The number of triples for each dataset}\label{lubm}
  \vspace{3mm}
  \begin{tabular}{cccccc}
    \hline
    dataset & LUBM\_10 & LUBM\_50 & LUBM\_100 & LUBM\_150 & LUBM\_200 \\
    \hline
    triples & 1449832 & 6822632 & 12297500 & 20652819 & 27643644 \\
    \hline
  \end{tabular}
\end{table}

\subsection{Experiment Performance}
We evaluate the performance of our method with KP and Cichlid, where KP adopts the executable strategy in \cite{je2015scalable}. All experiments run three times and the average value is listed as follow.

\begin{figure}[H]
\setlength{\abovecaptionskip}{0.25cm}
\setlength{\belowcaptionskip}{-0.65cm}
\centering
\begin{minipage}[t]{0.5\linewidth}
    \begin{tikzpicture}[scale=0.7]
     \footnotesize
      \begin{axis}[
    xlabel=$lubm/the\hspace{1mm} number\hspace{1mm} of\hspace{1mm} university$,
    ylabel=$reasoning\hspace{1mm}time/sec$,
    xtick={10, 50, 100, 150, 200},
    ytick={0, 1000, 2000, 3000, 4000, 5000, 6000, 7000,8000,9000},
    legend pos=north west,
    ymajorgrids=true,
    grid style=dashed,
]
\addplot[
    color=blue,
    mark=*,
    ]
   plot coordinates {
    (10, 460)(50, 1413)(100, 2460)(150, 4224)(200, 7080)};
\addlegendentry{RORS}
\addplot[
    color=red,
    mark=square,
    ]
    plot coordinates{(10, 480)(50, 1910)(100, 3250)(150, 5460)(200, 8850)};
\addlegendentry{KP}
\end{axis}
\end{tikzpicture}
\caption{The runtime of RORS and KP}\label{fig:time}
\end{minipage}%
\begin{minipage}[t]{0.5\linewidth}
\begin{tikzpicture}[scale=0.7]
  \begin{axis}[
   ylabel=$inferred \hspace{1mm} triples$,
	x tick label style={
		/pgf/number format/1000 sep=},
	enlargelimits=0.05,
	legend style={at={(0.5, -0.1)},
	anchor=north, legend columns=-1},
	ybar interval=0.7,
]
\addplot
	coordinates {(10, 38425.335) (50, 67471.298)
		 (100, 83201.436) (150, 86510.952) (200, 72749.017)};
\addplot
	coordinates {(10, 41840.289) (50, 60950.250)
		(100, 69645.680) (150, 75657.241) (200, 59880.319)};
\legend{RORS, Cichild}
\end{axis}
\end{tikzpicture}
\caption{The number of triples per sec}\label{fig:expres}
 \end{minipage}%
\end{figure}

Figure \ref{fig:time} displays the reasoning time of our approach and KP with different scale of data sets. The reasoning time includes the time of dividing input data and eliminating
duplicated triples. We can see that the reasoning time of RORS and KP increase almost linearly with the growth of data size. The result shows that our approach is better than the KP under the same environment. The performance of reasoning is improved by 30\% approximately.

Figure \ref{fig:expres} shows the number of inferred triples per second. Our method can infer more implicit triples than Cichlid and the performance is improved by 26\% approximately.
%There are a few reasons for the performance improved. Firstly, our method adopt more rules than Cichlid in OWL reasoning. For example, we add equivalent class and property rules into schema reasoning. And the rule O6 also derives many new triples. Secondly, the executable strategy used by our method is more efficient than Cichlid.
\section{Discussions}\label{sec:dis}
In this paper, we present an approach to enhancing the performance of the rule-based OWL reasoning on Spark based on a locally optimal executable strategy. Our method performs better than KP in reasoning strategy. Although the approach does't find the
optimal executable strategy for global rules, our method can be used as the valuable foundation for future research in the rule-based OWL reasoning. Therefore, in the future work, we plan to design some algorithms to find the optimal strategy for global rules.

There are many works to develop OWL reasoning systems, including early works, such as Pellet \cite{sirin2007pellet}, Jena \cite{mcbride2002jena}, and Sesame \cite{broekstra2002sesame}, etc. These reasoners use a composition
tree model for ontology to infer implicit information and exhibit both large time and space complexity. Due to the limitation of
computing resource and running speed, these systems can hardly achieve excellent performance. Therefore, many distributed reasoning
system emerged. In \cite{muhleisen2012large} and \cite{oren2009marvin}, they proposed a parallel reasoning method that the reasoning
rules are executed repeatedly until no extra data is generated. But there exists much more data communication cost. In \cite{weaver2009parallel},
Weaver and Handler proposed a data partitioning model based on MPI, but this method do not filter duplicate data.
\cite{urbani2009scalable} presented a distributed reasoning system which based on MapReduce. It analyzed the dependency between rules
and builded a dependence graph. But it generated large amount of useless middle data and huge data communication cost. Then Urbani
proposed a MapReduce-based parallel reasoning system with OWL-Horst rules called WebPIE \cite{urbani2010owl}. It can deal with large
scale ontology on a distributed computing cluster. However, WebPIE exhibits poor reasoning time. \cite{thakker2010pragmatic} are rule-based OWL reasoner. Although they can infer large scale triples, it costs too much reasoning time.

\cite{peters2013rule} proposed a rule-based reasoner that used massively parallel hardware to derive new facts based
on a given set of rules, but that implementation was limited by the size of processable input data as well as on the number of used
parallel hardware  devices. Seitz \cite{seitz2011rule} presented an OWL reasoner for embedded devices based on CLIPS. Urbani \cite{urbani2011querypie} proposed a hybrid rule-based reasoning method that combined forward and backward chaining, and
implemented a prototype named QueryPIE. Terminological triples are pre-computed before query, which is used to speed up backward-chaining at query time. In \cite{rong2015cichlid}, although the author has improved the reasoning time using Spark, it ignores the analysis of interdependence among the rules and does not give the optimal executable strategy.
Besides, MPPIE \cite{mppie} recently implemented the RDFS reasoning on Giraph.

\section{Acknowledgments}
This work is supported by the program of the National Natural Science
Foundation of China (NSFC) under 61502336, 61373035 and the
National High-tech R\&D Program of China (863 Program) under 2013AA013204.


\begin{thebibliography}{10}

\bibitem{broekstra2002sesame} Broekstra J., Kampman A., \& Van~Harmelen F.
    (2002)
\newblock Sesame: A generic architecture for storing and querying {RDF} and {RDF} schema.
\newblock In: {\em Proc. of ISWC 2002}, Springer, pp.~54--68.

\bibitem{Badder2001tableau} Baader F. \& Sattler U.(2001)
\newblock An overview of tableau algorithms for Description Logics.
\newblock {\em Studia Logica}, 69(1):5--40.

\bibitem{chang2011fuzzy} Chang L., Guilin Q., Haofen W. \& Yong Y. (2011)
\newblock Large scale fuzzy pD* reasoning using MapReduce.
\newblock In: {\em Proc. of ISWC 2011}, Springer, pp.~405--420.

\bibitem{rong2015cichlid} Gu R., Wang S., Wang F., Yuan C., \& Huang
    Y.(2015)
\newblock Cichlid: Efficient large scale {RDFS/OWL} reasoning with Spark.
\newblock In: {\em Proc. of IPDPS 2015}, IEEE, pp.~700--709.

\bibitem{guo2005lubm} Guo Y., Pan Z., \& Heflin J.(2005)
\newblock LUBM: A benchmark for {OWL} knowledge base systems.
\newblock {\em J. Web Sem.}, 3(2-3):158--182.


\bibitem{je2015scalable} Kim J. \& Park Y. (2015)
\newblock Scalable OWL-Horst ontology reasoning using {Spark}.
\newblock In: {\em Proc. of BigComp 2015}, pp.~79--86.

%\bibitem{kiryakov2005owlim} Kiryakov A., Ognyanov D., \& Manov D. (2005)
%\newblock {OWLIM} - {A} pragmatic semantic repository for {OWL}.
%\newblock In: {\em Proc. of WISE 2005}, pp.~182--192.

\bibitem{owl} L. McGuinness D., van Harmelen F. (2004)
\newblock Web Ontology Language.
\newblock {\em W3C Recommendation}.

\bibitem{mppie}
Lv X., Wang X., Feng Z., Rao G., Zhang X., \& Xu G.(2016)
\newblock MPPIE: RDFS parallel inference framework based on message passing ({\em in Chinese}). \newblock {\em Journal of Frontiers of Computer Science and Technology}, 10(4): 451-465.

\bibitem{mcbride2002jena} McBride B. (2002)
\newblock Jena: A semantic web toolkit.
\newblock {\em Internet computing 2002}, IEEE, 6(5):55--59.

\bibitem{muhleisen2012large} M{\"u}hleisen H. \& Dentler K. (2012)
\newblock Large-scale storage and reasoning for semantic data using swarms.
\newblock {\em IEEE Comp. Int. Mag.}, 7(2):32--44.

\bibitem{maeda2014mapreduce} Maeda R., Ohta N., \& Kuwabara K.(2014)
\newblock MapReduce-based implementation of a rule system.
\newblock In: {\em Recent Developments in Computational Collective Intelligence 2014}, Springer, pp.~197--206.

\bibitem{oren2009marvin} Oren E., Kotoulas S., Anadiotis G., Siebes R., ten
    Teije A., \& van Harmelen F.(2009)
\newblock Marvin: Distributed reasoning over large-scale semantic web data.
\newblock  {\em J. Web Sem.}, 7(4):305--316.

\bibitem{peters2013rule} Peters M., Brink C., Sachweh S., \& Z{\"{u}}ndorf
    A. (2013)
\newblock Rule-based reasoning on massively parallel hardware.
\newblock In: {\em Proc. of SSWS 2013 at ISWC}, pp.~33--49.

\bibitem{sirin2007pellet} Sirin E., Parsia B., C. Grau B., Kalyanpur A., \&
    Katz Y. (2007)
\newblock Pellet: A practical {OWL-DL} reasoner.
\newblock {\em J. Web Sem.}, 5(2):51--53.

\bibitem{seitz2011rule} Seitz C., \& Sch{\"{o}}nfelder R. (2011)
\newblock Rule-Based {OWL} reasoning for specific embedded devices.
\newblock In: {\em Proc. of ISWC 2011}, Springer, pp.~237-252.

\bibitem{thakker2010pragmatic} Thakker D., Osman T., Gohil S., \& Lakin
    P.(2010)
\newblock A pragmatic approach to semantic repositories benchmarking.
\newblock In: {\em Proc. of ESWC 2010}, Springer, pp.~379--393.

\bibitem{urbani2009scalable} Urbani J., Kotoulas S., Oren E., \&
    Van~Harmelen F. (2009)
\newblock Scalable distributed reasoning using mapreduce.
\newblock In: {\em Proc. of ISWC 2009}, Springer, pp.~634--649.

\bibitem{urbani2010owl} Urbani J., Kotoulas S., Maassen J., Van~Harmelen F.,
    \& Bal H.(2010)
\newblock OWL reasoning with {WebPIE}: Calculating the closure of 100 billion triples.
\newblock In: {\em Proc. of ESWC 2010}, Springer, vol.1, pp.~213--227.

\bibitem{urbani2011querypie} Urbani J., van Harmelen F., Schlobach S.,
    \& E. Bal H. (2011)
\newblock Query{PIE}: Backward reasoning for {OWL} Horst over very large knowledge bases.
\newblock In: {\em Proc. of ISWC 2011}, Springer, pp.~730--745.

\bibitem{weaver2009parallel} Weaver J. \& A. Hendler J. (2009)
\newblock Parallel materialization of the finite RDFS closure for hundreds of millions of triples.
\newblock In: {\em Proc. of ISWC 2009}, Springer, pp.~682--697.

\bibitem{wu2013distributed} Wu H., Liu J., Ye D., Zhong H., \& Wei J. (2013)
\newblock A distributed rule execution mechanism based on mapreduce in sematic web reasoning.
\newblock In: {\em Proc. of Internetware 2013}, ACM, Article No. 6.


\bibitem{spark}
Zaharia M., Chowdhury M., J. Franklin M., Shenker S., \& Stoica I. (2010)
\newblock Spark: Cluster computing with working sets.
\newblock In: {\em Proc. of HotCloud 2010}, Boston, MA, USA.

\bibitem{zaharia2012resilient} Zaharia M., Chowdhury M., Das T., Dave A., Ma
    J., McCauley M., J. Franklin M., Shenker S., \& Stoica I. (2012)
\newblock Resilient distributed datasets: A fault-tolerant abstraction for in-memory cluster computing.
\newblock In: {\em Proc. of NSDI 2012 at USENIX}, pp.~15--28.

\end{thebibliography}
\end{document}